\documentclass[journal,10pt,twocolumn]{IEEEtran}
\usepackage{epsfig}
\usepackage{graphicx}
\usepackage{subfigure}

\newtheorem{proposition}{Proposition}

\usepackage{amssymb}
\usepackage{makeidx}
\usepackage{amsmath}
\usepackage{graphicx}
\usepackage{epsf}
\usepackage{epsfig}
\usepackage{psfig}
\usepackage{ccaption}
\usepackage{array}
\usepackage{tabularx}
\usepackage{multirow}
\usepackage{epsfig}
\usepackage{cite}
\begin{document}
%

\title{Multi-agent Q-Learning of Channel Selection in Multi-user Cognitive Radio Systems:\\ A Two by Two Case}
%
\author{Husheng Li
\thanks{H. Li is with the Department of Electrical Engineering and Computer Science, the University of Tennessee, Knoxville, TN, 37996 (email:
husheng@eecs.utk.edu). This work was supported by the National
Science Foundation under grant CCF-0830451.}}

\maketitle
\begin{abstract}
Resource allocation is an important issue in cognitive radio
systems. It can be done by carrying out negotiation among secondary
users. However, significant overhead may be incurred by the
negotiation since the negotiation needs to be done frequently due to
the rapid change of primary users' activity. In this paper, a
channel selection scheme without negotiation is considered for
multi-user and multi-channel cognitive radio systems. To avoid
collision incurred by non-coordination, each user secondary learns
how to select channels according to its experience. Multi-agent
reinforcement leaning (MARL) is applied in the framework of
Q-learning by considering the opponent secondary users as a part of
the environment. The dynamics of the Q-learning are illustrated
using Metrick-Polak plot. A rigorous proof of the convergence of
Q-learning is provided via the similarity between the Q-learning and
Robinson-Monro algorithm, as well as the analysis of convergence of
the corresponding ordinary differential equation (via Lyapunov
function). Examples are illustrated and the performance of learning
is evaluated by numerical simulations.
\end{abstract}
\section{Introduction}
In recent years, cognitive radio has attracted extensive studies in
the community of wireless communications. It allows users without
license (called secondary users) to access licensed frequency bands
when the licensed users (called primary users) are not present.
Therefore, the cognitive radio technique can substantially alleviate
the problem of under-utilization of frequency spectrum
\cite{Mitola}\cite{Mitola1999}.

The following two problems are key to the cognitive radio systems:
\begin{itemize}
\item Resource mining, i.e. how to detect the available resource (the frequency bands that are not being used by
primary users); usually it is done by carrying out spectrum sensing.

\item Resource allocation, i.e. how to allocate the detected available
resource to different secondary users.
\end{itemize}

Substantial work has been done for the resource mining. Many signal
processing techniques have been applied to sense the frequency
spectrum \cite{Zhao2007}, e.g. cyclostationary feature
\cite{Kim2007}, quickest change detection \cite{Li2008},
collaborative spectrum sensing \cite{Ghasemi2005}. Meanwhile, plenty
of researches have been conducted for the resource allocation in
cognitive radio systems \cite{Nyyato} \cite{Kloeck2006}. Typically,
it is assumed that the secondary users exchange information about
detected available spectrum resources and then negotiate the
resource allocation according to their own requirements of traffic
(since the same resource cannot be shared by different secondary
users if orthogonal transmission is assumed). These studies
typically apply theories in economics, e.g. game theory, bargaining
theory or microeconomics.

However, in many applications of cognitive radio, such a negotiation
based resource allocation may incur significant overhead. In
traditional wireless communication systems, the available resource
is almost fixed (even if we consider the fluctuation of channel
quality, the change of available resource is still very slow and
thus can be considered stationary). Therefore, the negotiation need
not be carried out frequently and the negotiation result can be
applied for a long period of data communication, thus incurring
tolerable overhead. However, in many cognitive radio systems, the
resource may change very rapidly since the activity of primary users
may be highly dynamic. Therefore, the available resource needs to be
updated very frequently and the data communication period should be
fairly short since minimum violation to primary users should be
guaranteed. In such a situation, the negotiation of resource
allocation may be highly inefficient since a substantial portion of
time needs to be used for the negotiation. To alleviate such an
inefficiency, high speed transceivers need to be used to minimize
the time consumed on negotiation. Particularly, the turn-around
time, i.e. the time needed to switch from receiving (transmitting)
to transmitting (receiving) should be very small, which is a
substantial challenge to hardware design.

\begin{figure}
  \centering
  \includegraphics[scale=0.5]{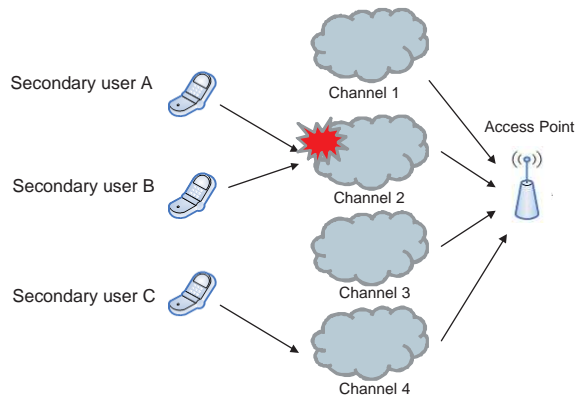}
  \caption{Illustration of competition and conflict in multi-user and multi-channel cognitive radio systems.}\label{fig:compete}
\end{figure}

Motivated by the previous discussion and observation, in this paper,
we study the problem of spectrum access without negotiation in
multi-user and multi-channel cognitive radio systems. In such a
scheme, each secondary user senses channels and then choose an idle
frequency channel to transmit data, as if no other secondary user
exists. If two secondary users choose the same channel for data
transmission, they will collide with each other and the data packets
cannot be decoded by the receiver(s). Such a procedure is
illustrated in Fig. \ref{fig:compete}, where three secondary users
access an access point via four channels. Since there is no mutual
communication among these secondary users, conflict is unavoidable.
However, the secondary users can try to learn how to avoid each
other, as well as channel qualities (we assume that the secondary
users have no \textit{a priori} information about the channel
qualities), according to its experience. In such a context, the
cognition procedure includes not only the frequency spectrum but
also the behavior of other secondary users.

To accomplish the task of learning channel selection, multi-agent
reinforcement learning (MARL) \cite{Bysoniu2008} is a powerful tool.
One challenge of MARL in our context is that the secondary users do
not know the payoffs (thus the strategy) of each other in each
stage; thus the environment of each secondary user, including its
opponents, is dynamic and may not assure convergence of learning. In
such a situation, fictitious play
\cite{Fudenberg1998}\cite{Robinson1969}, which estimates other
users' strategy and plays the best response, can assure convergence
to a Nash equilibrium point within certain assumptions. As an
alternative way, we adopt the principle of Q-learning, i.e.
evaluating the values of different actions in an incremental way.
For simplicity, we consider only the case of two secondary users and
two channels. By applying the theory of stochastic approximation
\cite{Kushner2003}, we will prove the main result of this paper,
i.e. the learning converges to a stationary point regardless of the
initial strategies (Propositions \ref{prop:eqi_ODE} and
\ref{prop:conv_ODE} ). Note that our study is one extreme of the
resource allocation problem since no negotiation is considered while
the other extreme is full negotiation to achieve optimal
performance. It is interesting to study the intermediate case, i.e.
limited negotiation for resource allocation. However, it is beyond
the scope of this paper.

The remainder of this paper is organized as follows. In Section
\ref{sec:model}, the system model is introduced. The proposed
Q-learning for channel selection is explained in Section
\ref{sec:Q}. Intuitive explanation and rigorous proof for
convergence are explained in Sections \ref{sec:intuition} and
\ref{sec:sto_approx}, respectively. The numerical results are
provided in Section \ref{sec:numerical} while the conclusions are
drawn in Section \ref{sec:conclusion}.

\section{System Model}\label{sec:model}
For simplicity, we consider only two secondary users, denoted by $A$
and $B$, and two channels, denoted by 1 and 2. The reward to
secondary user $i$, $i=A,B$, of channel $j$, $j=1,2$, is $R_{ij}$ if
secondary user transmits data over channel $j$ and channel $j$ is
not interrupted by primary user or the other secondary user;
otherwise the reward is 0 since the secondary user cannot convey any
information over this channel. For simplicity, we denote by $j^-$
the other user (channel) different from user (channel) $j$.

The following assumptions are placed throughout this paper.
\begin{itemize}
\item The rewards $\{R_{ij}\}$ are unknown to both secondary users.
They are fixed throughout the game.

\item Both secondary users can sense both channels simultaneously, but can choose
only one channel for data transmission. It is more interesting and
challenging to study the case that the secondary users can sense
only one channel, thus forming a partially observable game. However,
it is beyond the scope of this paper.

\item We consider only the case that both channels are available
since the actions that the secondary users can take are obvious
(transmit over the only available channel or not transmit if no
channel is available). Thus, we ignore the task of sensing the
frequency spectrum, which has been well studied by many researchers,
and focus on only the cognition of the other secondary user's
behavior.

\item There is no communication between the two secondary users.
\end{itemize}

\section{Game and $Q$-learning}\label{sec:Q}
In this section, we introduce the corresponding game and the
application of Q-learning to the channel selection problem.

\subsection{Game of Channel Selection}
\begin{figure}
  \centering
  \includegraphics[scale=0.5]{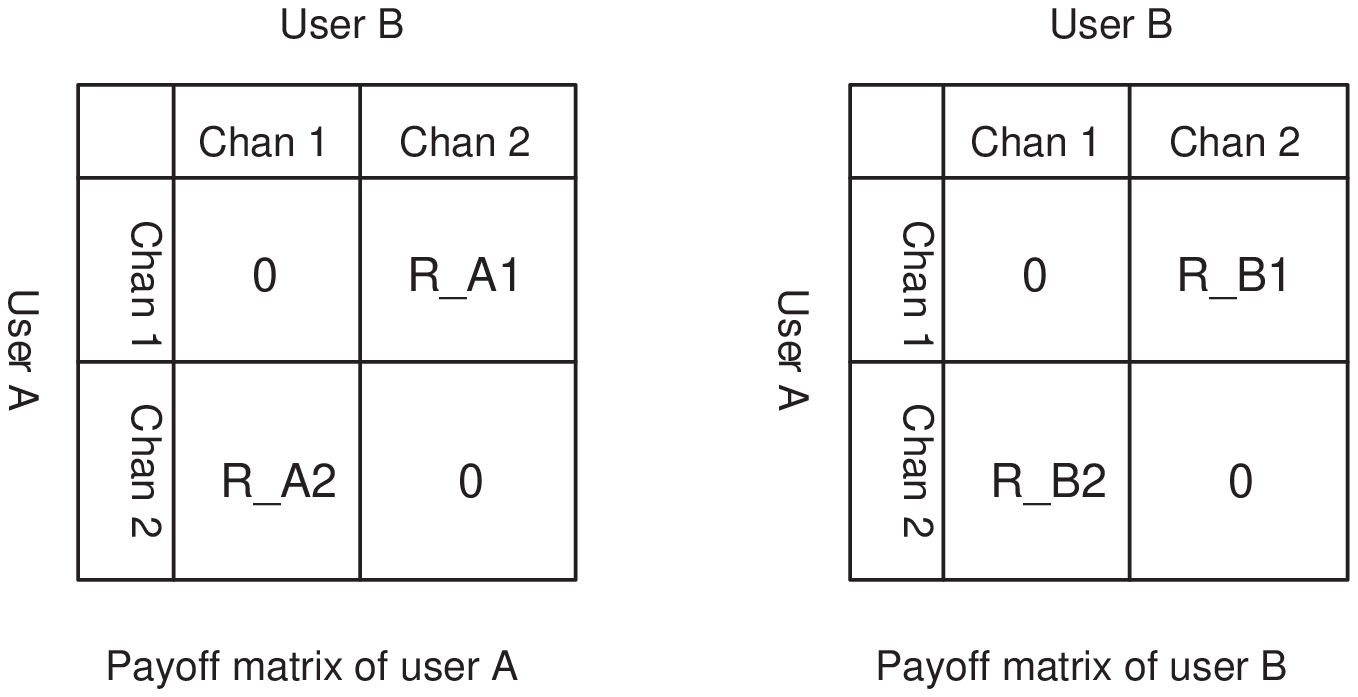}
  \caption{Payoff matrices in the game of channel selection.}\label{fig:game}
\end{figure}

The channel selection problem is a $2\times 2$ game, in which the
payoff matrices are given in Fig. \ref{fig:game}. Note that the
actions, denoted by $a_i(t)$ for user $i$ at time $t$, in the game
are the selections of channels. Obviously, the diagonal elements in
the payoff matrices are all zero since conflict incurs zero reward.

It is easy to verify that there are two Nash equilibrium points in
the game, i.e. the strategies such that unilaterally changing
strategy incurs its own performance degradation. Both equilibrium
points are pure, i.e. $a_A=1,a_B=2$ and $a_A=2,a_B=1$ (orthogonal
transmission).

\subsection{$Q$-function}
Since we assume that both channels are available, then there is only
one state in the system. Therefore, the $Q$-function is simply the
expected reward of each action (note that, in traditional learning
in stochastic environment, the $Q$-function is defined over the pair
of state and action), i.e.
\begin{eqnarray}
Q(a)=E[R(a)],
\end{eqnarray}
where $a$ is the action, $R$ is the reward dependent on the action
and the expectation is over the randomness of the other user's
action. Since the action is the selection of channel, we denote by
$Q_{ij}$ the value of selecting channel $j$ by secondary user $i$.

\subsection{Exploration}
In contrast to fictitious play \cite{Fudenberg1998}, which is
deterministic, the action in $Q$-learning is stochastic to assure
that all actions will be tested. We consider Boltzmann distribution
for random exploration, i.e.
\begin{eqnarray}\label{eq:exploration}
P(\mbox{user $i$ choose channel
$j$})=\frac{e^{Q_{ij}/\gamma}}{e^{Q_{ij}/\gamma}+e^{Q_{ij^-}/\gamma}},
\end{eqnarray}
where $\gamma$ is called temperature, which controls the frequency
of exploration.

Obviously, when secondary user $i$ selects channel $j$, the expected
reward is given by
\begin{eqnarray}
E\left[R_i(j)\right]=\frac{R_{ij}e^{Q_{i^-j^-}/\gamma}}{e^{Q_{i^-j}/\gamma}+e^{Q_{i^-j^-}/\gamma}},
\end{eqnarray}
since secondary user $i^-$ chooses channel $j$ with probability
$\frac{e^{Q_{i^-j}/\gamma}}{e^{Q_{i^-j}/\gamma}+e^{Q_{i^-j^-}/\gamma}}$
(collision happens and secondary user $i$ receives no reward) and
channel $j^-$ with probability
$\frac{e^{Q_{i^-j^-}/\gamma}}{e^{Q_{i^-j}/\gamma}+e^{Q_{i^-j^-}/\gamma}}$
(the transmissions are orthogonal and secondary user $i$ receives
reward $R_{ij}$).

\subsection{Updating $Q$-Functions}
In the procedure of $Q$-learning, the $Q$-functions are updated
after each spectrum access via the following rule:
\begin{eqnarray}\label{eq:Q_learning}
Q_{ij}(t+1)=(1-\alpha_{ij})Q_{ij}(t)+\alpha_{ij}(t)r_{i}(t)I(a_i(t)=j),
\end{eqnarray}
where $\alpha_{ij}(t)$ is a step factor (when channel $j$ is not
selected by user $i$, $\alpha_{ij}(t)=0$) and $r_i(t)$ is the reward
of secondary user $i$ and $I$ is characteristic function for the
event that channel $j$ is selected at the $t$-th spectrum access.
Our study is focused on the dynamics of (\ref{eq:Q_learning}). To
assure convergence, we assume that
\begin{eqnarray}
\sum_{t=1}^\infty \alpha_{ij}(t)=\infty,\qquad \forall i=A,B,j=1,2.
\end{eqnarray}

\section{Intuition on Convergence}\label{sec:intuition}
\begin{figure}
  \centering
  \includegraphics[scale=0.5]{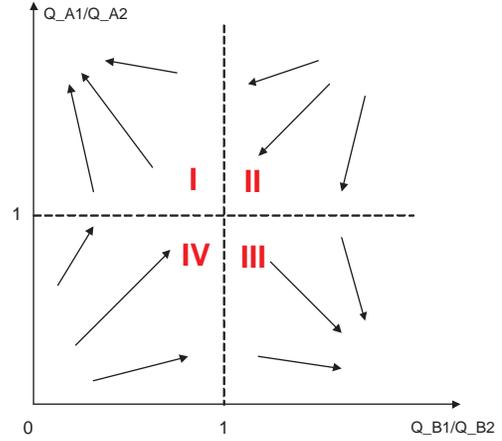}
  \caption{Illustration of the dynamics in the Q-learning.}\label{fig:graph}
\end{figure}

As will be shown in Propositions \ref{prop:eqi_ODE} and
\ref{prop:conv_ODE}, the updating rule of $Q$ functions in
(\ref{eq:Q_learning}) will converge to a stationary equilibrium
point close to Nash equilibrium if the step factor satisfies certain
conditions. Before the rigorous proof, we provide an intuitive
explanation for the convergence using the geometric argument
proposed in \cite{Metrick1994}.

The intuitive explanation is provided in Fig. \ref{fig:graph} (we
call it \textit{Metrick-Polak plot} since it was originally proposed
by A. Metrick and B. Polak in \cite{Metrick1994}), where the axises
are $\mu_A=\frac{Q_{A1}}{Q_{A2}}$ and $\mu_B=\frac{Q_{B1}}{Q_{B2}}$,
respectively. As labeled in the figure, the plane is divided into
four regions by two lines $\mu_A=1$ and $\mu_B=1$, in which the
dynamics of $Q$-learning are different. We discuss these four
regions separately:
\begin{itemize}
\item Region I: in this region, $Q_{A1}>Q_{A2}$; therefore,
secondary user $A$ prefers visiting channel 1; meanwhile, secondary
user $B$ prefers accessing channel 2 since $Q_{B1}>Q_{B2}$; then,
with large probability, the strategies will converge to the Nash
equilibrium point in which secondary users $A$ and $B$ access
channels 1 and 2, respectively.

\item Region II: in this region, both secondary users prefer
accessing channel 1, thus causing many collisions. Therefore, both
$Q_{A1}$ and $Q_{B1}$ will be reduced until entering either region I
or region III.

\item Region III: similar to region I.

\item Region IV: similar to region II.
\end{itemize}

Then, we observe that the points in Regions II and IV are unstable
and will move into Region I or III with large probability. In
Regions I and III, the strategy will move close to the Nash
equilibrium points with large probability. Therefore, regardless
where the initial point is, the updating rule in
(\ref{eq:Q_learning}) will generate a stationary equilibrium point
with large probability.

\section{Stochastic Approximation Based
Convergence}\label{sec:sto_approx}

In this section, we prove the convergence of the $Q$-learning.
First, we find the equivalence between the updating rule
(\ref{eq:Q_learning}) and Robbins-Monro iteration \cite{Robbins1951}
for solving an equation with unknown expression. Then, we apply the
conclusion in stochastic approximation \cite{Kushner2003} to relate
the dynamics of the updating rule to an ordinary differential
equation (ODE) and prove the stability of the ODE.

\subsection{Robbins-Monro Iteration}
At a stationary point, the expected values of $Q$-functions satisfy
the following four equations:
\begin{eqnarray}\label{eq:equil_equa0}
Q_{ij}=\frac{R_{ij}e^{Q_{i^{-}j^{-}}/\gamma}}{e^{Q_{i^{-}j}/\gamma}+e^{Q_{i^{-}j^-}/\gamma}},\qquad
i=A,B,\mbox{  }j=1,2.
\end{eqnarray}

Define $\mathbf{q}=\left(Q_{A1},Q_{A2},Q_{B1},Q_{B2}\right)^T$. Then
(\ref{eq:equil_equa0}) can be rewritten as
\begin{eqnarray}\label{eq:equil_equa}
\mathbf{g}(\mathbf{q})=\mathbf{A}(\mathbf{q})\mathbf{r}-\mathbf{q}=0,
\end{eqnarray}
where $\mathbf{r}=\left(R_{A1},R_{A2},R_{B1},R_{B2}\right)^T$ and
the matrix $\mathbf{A}$ (as a function of $\mathbf{q}$) is given by
\begin{eqnarray}
\mathbf{A}_{ij}= \left\{
\begin{array}{ll}
\frac{e^{Q_{i^-j^-}/\gamma}}{e^{Q_{i^-j}/\gamma}+e^{Q_{i^-j^-}/\gamma}},&\mbox{ if }i=j\\
0,&\mbox{ if }i\neq j
\end{array}
\right..
\end{eqnarray}

Then, the updating rule in (\ref{eq:Q_learning}) is equivalent to
solving the equation (\ref{eq:equil_equa}) (the expression of the
equation is unknown since the rewards, as well as the strategy of
the other user, are unknown) using Robbins-Monro algorithm
\cite{Kushner2003}, i.e.
\begin{eqnarray}
\mathbf{q}(t+1)=\mathbf{q}(t)+\alpha(t)\mathbf{Y}(t),
\end{eqnarray}
where $\mathbf{Y}(t)$ is a random observation on function $g$
contaminated by noise, i.e.
\begin{eqnarray}\label{eq:stoch_approx}
\mathbf{Y}(t)&=&\mathbf{{r}}(t)-\mathbf{q}(t)\nonumber\\
&=&\bar{\mathbf{r}}(t)-\mathbf{q}(t)+\mathbf{\hat{r}}(t)-\bar{\mathbf{r}}\nonumber\\
&=&\mathbf{g}_t(\mathbf{q}(t))+\delta M(t),
\end{eqnarray}
where $\mathbf{g}_t(\mathbf{q}(t))=\bar{\mathbf{r}}-\mathbf{q}(t)$,
$\delta M(t)=\mathbf{\hat{r}}(t)-\mathbf{q}^*$ is noise and (recall
that $r_{i}(t)$ means the reward of secondary user $i$ at time $t$)
\begin{eqnarray}
\mathbf{\bar{r}}(t)=\mathbf{A}(\mathbf{q}(t))\mathbf{r}.
\end{eqnarray}

\subsection{ODE and Convergence}
The procedure of using Robbins-Monro algorithm (i.e. the updating of
$Q$-function) is the stochastic approximation of the solution of the
equation. It is well known that the convergence of such a procedure
can be characterized by an ODE. Since the noise $\delta M(t)$ in
(\ref{eq:stoch_approx}) is a Martingale difference, we can verify
the conditions in Theorem 12.3.5 in \cite{Kushner2003} (the
verification is omitted due to limited length of this paper) and
obtain the following proposition:
\begin{proposition}\label{prop:eqi_ODE}
With probability 1, the sequence $\mathbf{q}(t)$ converges to some
limit set of the ODE
\begin{eqnarray}\label{eq:ODE}
\dot{\mathbf{q}}=\mathbf{g}(\mathbf{q}).
\end{eqnarray}
\end{proposition}

What remains to do is to analyze the convergence property of the ODE
(\ref{eq:ODE}). We obtain the following proposition:
\begin{proposition}\label{prop:conv_ODE}
The solution of ODE (\ref{eq:ODE}) converges to the stationary point
determined by (\ref{eq:equil_equa}).
\end{proposition}
\begin{proof}
We apply Lyapunov's method to analyze the convergence of the ODE
(\ref{eq:ODE}). We define the Lyapunov function as
\begin{eqnarray}
V(t)&=&\|\mathbf{g}(t)\|^2\nonumber\\
 &=&\sum \left(\bar{r}_{ij}(t)-Q_{ij}(t)\right)^2.
\end{eqnarray}

Then, we examine the derivative of the Lyapunov function with
respect to time $t$, i.e.
\begin{eqnarray}
\frac{dV(t)}{dt}&=&2\sum\frac{d(\bar{r}_{ij}(t)-Q_{ij}(t))}{dt}\left(\hat{r}_{ij}(t)-Q_{ij}(t)\right)\nonumber\\
&=&2\sum\frac{d \epsilon_{ij}(t)}{dt}\epsilon_{ij}(t),
\end{eqnarray}
where $\epsilon_{ij}(t)\triangleq \hat{r}_{ij}(t)-Q_{ij}(t)$.

We have
\begin{eqnarray}
\frac{d\epsilon_{ij}(t)}{dt}&=&\frac{d \bar{r}_{ij}(t)}{dt}-\frac{d
Q_{ij}(t)}{dt}\nonumber\\
&=&\frac{d \bar{r}_{ij}(t)}{dt}-\epsilon_{ij}(t),
\end{eqnarray}
where $r_{ij}(t)=(\mathbf{A}\mathbf{r})_{ij}$ and we applied the ODE
(\ref{eq:ODE}).

Then, we focus on the computation of $\frac{d r_{ij}(t)}{dt}$. When
$i=A$ and $j=1$, we have
\begin{eqnarray}
\frac{d
\bar{r}_{A1}(t)}{dt}&=&\frac{d}{dt}\left(\frac{R_{A1}e^{Q_{B2}/\gamma}}{e^{Q_{B1}/\gamma}+e^{Q_{B2}/\gamma}}\right)\nonumber\\
&=&\frac{R_{A1}e^{Q_{B1}/\gamma}e^{Q_{B2}/\gamma}}{\gamma\left(e^{Q_{B1}/\gamma}+e^{Q_{B2}/\gamma}\right)^2}\nonumber\\
&\times&\left(\frac{d Q_{B2}(t)}{dt}-\frac{d
Q_{B1}(t)}{dt}\right)\nonumber\\
&=&\frac{R_{A1}e^{Q_{B1}/\gamma}e^{Q_{B2}/\gamma}}{\gamma\left(e^{Q_{B1}/\gamma}+e^{Q_{B2}/\gamma}\right)^2}\nonumber\\
&\times&\left(\epsilon_{B2}(t)-\epsilon_{B1}(t)\right),
\end{eqnarray}
where we applied the ODE (\ref{eq:ODE}) again.

Using similar arguments, we have
\begin{eqnarray}
\frac{d \bar{r}_{A2}(t)}{dt}
&=&\frac{R_{A2}e^{Q_{B1}/\gamma}e^{Q_{B2}/\gamma}}{\gamma\left(e^{Q_{B1}/\gamma}+e^{Q_{B2}/\gamma}\right)^2}\nonumber\\
&\times&\left(\epsilon_{B1}(t)-\epsilon_{B2}(t)\right),
\end{eqnarray}
and
\begin{eqnarray}
\frac{d \bar{r}_{B1}(t)}{dt}
&=&\frac{R_{B1}e^{Q_{A1}/\gamma}e^{Q_{A2}/\gamma}}{\gamma\left(e^{Q_{A1}/\gamma}+e^{Q_{A2}/\gamma}\right)^2}\nonumber\\
&\times&\left(\epsilon_{A2}(t)-\epsilon_{A1}(t)\right),
\end{eqnarray}
and
\begin{eqnarray}
\frac{d \bar{r}_{B2}(t)}{dt}
&=&\frac{R_{B2}e^{Q_{A1}/\gamma}e^{Q_{A2}/\gamma}}{\gamma\left(e^{Q_{A1}/\gamma}+e^{Q_{A2}/\gamma}\right)^2}\nonumber\\
&\times&\left(\epsilon_{A1}(t)-\epsilon_{A2}(t)\right).
\end{eqnarray}

Combining the above results, we have
\begin{eqnarray}
\frac{1}{2}\frac{dV(t)}{dt}&=&-\sum \epsilon_{ij}^2(t)\nonumber\\
&+&C_{12}\epsilon_{A1}\epsilon_{B2}-C_{11}\epsilon_{A1}\epsilon_{B1}\nonumber\\
&+&C_{21}\epsilon_{A2}\epsilon_{B1}-C_{22}\epsilon_{A2}\epsilon_{B2}
\end{eqnarray}
where
\begin{small}
\begin{eqnarray}
C_{12}=\left(\frac{R_{A1}e^{Q_{B1}/\gamma}e^{Q_{B2}/\gamma}}{\gamma\left(e^{Q_{B1}/\gamma}+e^{Q_{B2}/\gamma}\right)^2}+\frac{R_{B2}e^{Q_{A1}/\gamma}e^{Q_{A2}/\gamma}}{\gamma\left(e^{Q_{A1}/\gamma}+e^{Q_{A2}/\gamma}\right)^2}\right),
\end{eqnarray}
\end{small}
and
\begin{small}
\begin{eqnarray}
C_{11}=\left(\frac{R_{A1}e^{Q_{B1}/\gamma}e^{Q_{B2}/\gamma}}{\gamma\left(e^{Q_{B1}/\gamma}+e^{Q_{B2}/\gamma}\right)^2}+\frac{R_{B1}e^{Q_{A1}/\gamma}e^{Q_{A2}/\gamma}}{\gamma\left(e^{Q_{A1}/\gamma}+e^{Q_{A2}/\gamma}\right)^2}\right),
\end{eqnarray}
\end{small}
and
\begin{small}
\begin{eqnarray}
C_{21}=\left(\frac{R_{A2}e^{Q_{B1}/\gamma}e^{Q_{B2}/\gamma}}{\gamma\left(e^{Q_{B1}/\gamma}+e^{Q_{B2}/\gamma}\right)^2}+\frac{R_{B1}e^{Q_{A1}/\gamma}e^{Q_{A2}/\gamma}}{\gamma\left(e^{Q_{A1}/\gamma}+e^{Q_{A2}/\gamma}\right)^2}\right),
\end{eqnarray}
\end{small}
and
\begin{small}
\begin{eqnarray}
C_{22}=\left(\frac{R_{A2}e^{Q_{B1}/\gamma}e^{Q_{B2}/\gamma}}{\gamma\left(e^{Q_{B1}/\gamma}+e^{Q_{B2}/\gamma}\right)^2}+\frac{R_{B2}e^{Q_{A1}/\gamma}e^{Q_{A2}/\gamma}}{\gamma\left(e^{Q_{A1}/\gamma}+e^{Q_{A2}/\gamma}\right)^2}\right),
\end{eqnarray}
\end{small}

It is easy to verify that
\begin{eqnarray}
\frac{e^{Q_{B1}/\gamma}e^{Q_{B2}/\gamma}}{\left(e^{Q_{B1}/\gamma}+e^{Q_{B2}/\gamma}\right)^2}<1.
\end{eqnarray}

Now, we assume that $R_{ij}< 2\gamma$, then $C_{ij}< 2$. Therefore,
we have
\begin{eqnarray}
\frac{1}{2}\frac{dV(t)}{dt}&<& -\sum
\epsilon_{ij}^2(t)+2\epsilon_{A1}\epsilon_{B2}+2\epsilon_{A2}\epsilon_{B1}\nonumber\\
&=&-(\epsilon_{A1}-\epsilon_{B2})^2-(\epsilon_{A2}-\epsilon_{B1})^2<0.
\end{eqnarray}
Therefore, when $R_{ij}< 2\gamma$, the derivative of the Lyapunov
function is strictly negative, which implies that the ODE
(\ref{eq:ODE}) converges to a stationary point.

The final step of the proof is to remove the condition $R_{ij}<
2\gamma$. This is straightforward since we notice that the
convergence is independent of the scale of the reward $R_{ij}$.
Therefore, we can always scale the reward such that $R_{ij}<
2\gamma$. This concludes the proof.

\end{proof}

\section{Numerical Results}\label{sec:numerical}
In this section, we use numerical simulations to demonstrate the
theoretical results obtained in previous sections. For all
simulations, we use $\alpha_{ij}(t)=\frac{\alpha_0}{t}$, where
$\alpha_0$ is the initial learning factor.

\subsection{Dynamics}
\begin{figure}
  \centering
  \includegraphics[scale=0.4]{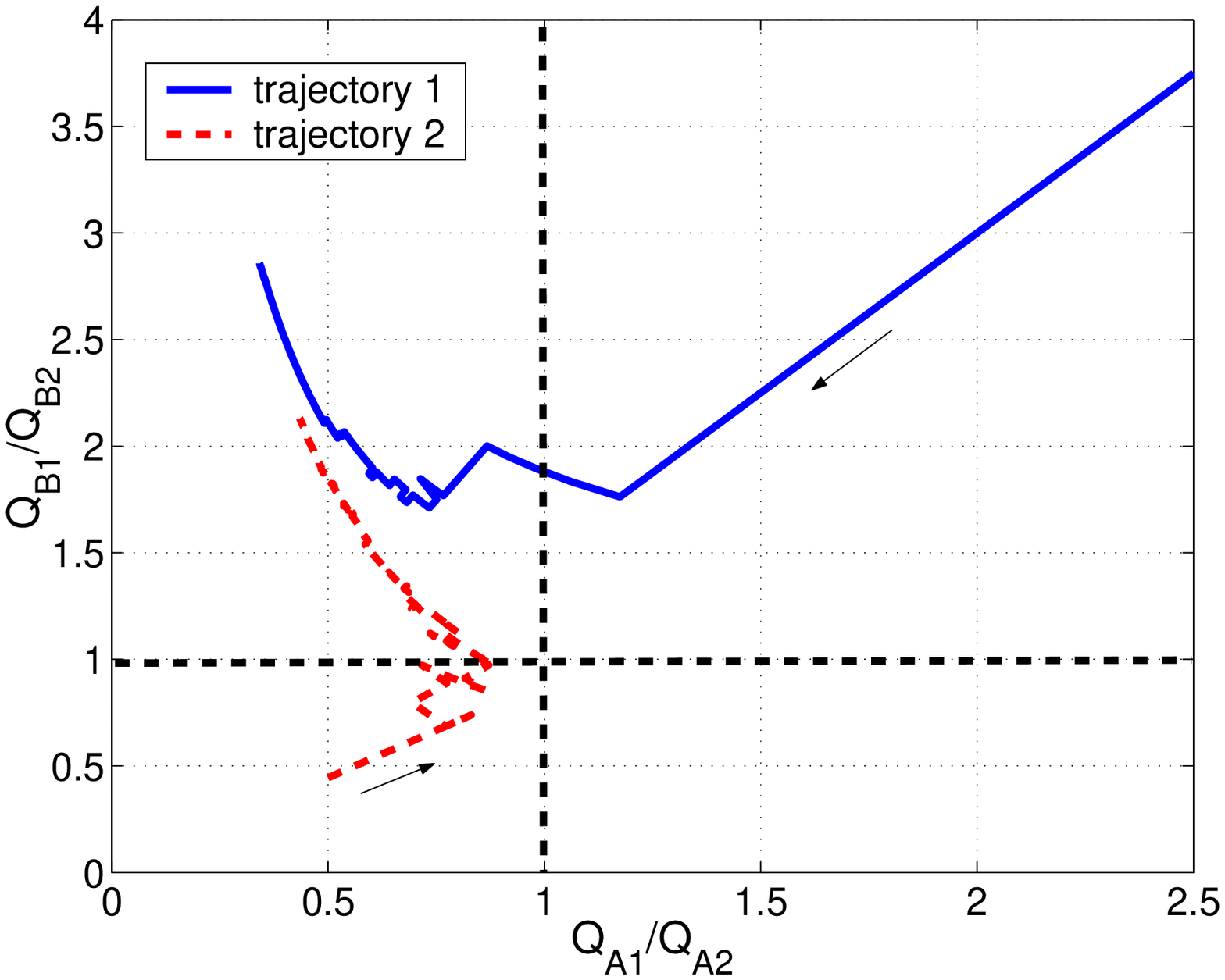}
  \caption{An example of dynamics of the $Q$-learning.}\label{fig:dynamics}
\end{figure}

\begin{figure}
  \centering
  \includegraphics[scale=0.4]{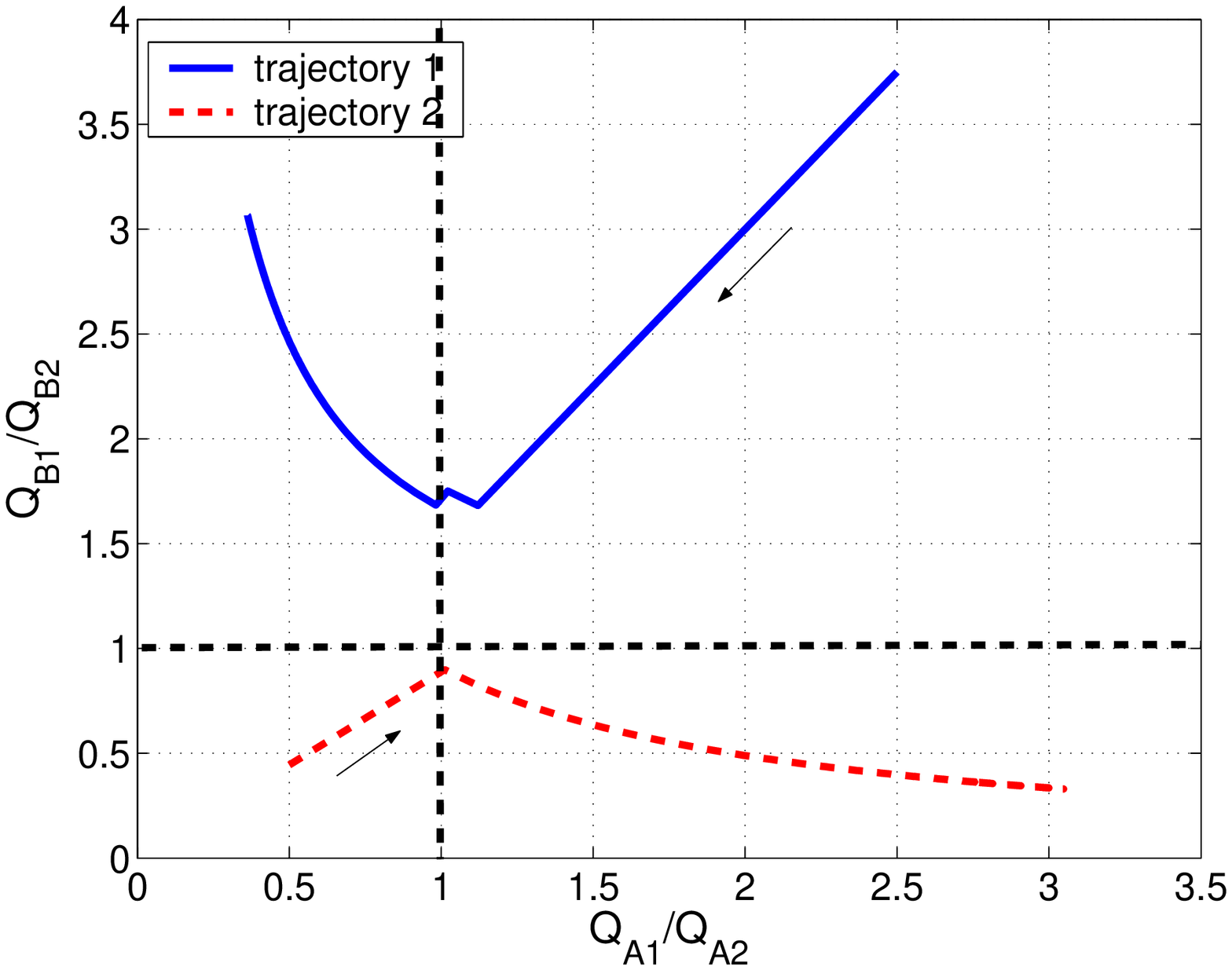}
  \caption{An example of dynamics of the $Q$-learning.}\label{fig:dynamics2}
\end{figure}

Figures \ref{fig:dynamics} and \ref{fig:dynamics2} show the dynamics
of $\frac{Q_{A1}}{Q_{A2}}$ versus $\frac{Q_{B1}}{Q_{B2}}$ of several
typical trajectories. Note that $\gamma=0.1$ in Fig.
\ref{fig:dynamics} and $\gamma=0.01$ in Fig. \ref{fig:dynamics2}. We
observe that the trajectories move from unstable regions (II and IV
in Fig. \ref{fig:graph}) to stable regions (I and III in Fig.
\ref{fig:graph}). We also observe that the trajectories for smaller
temperature $\gamma$ is smoother since less explorations are carried
out.

\begin{figure}
  \centering
  \includegraphics[scale=0.4]{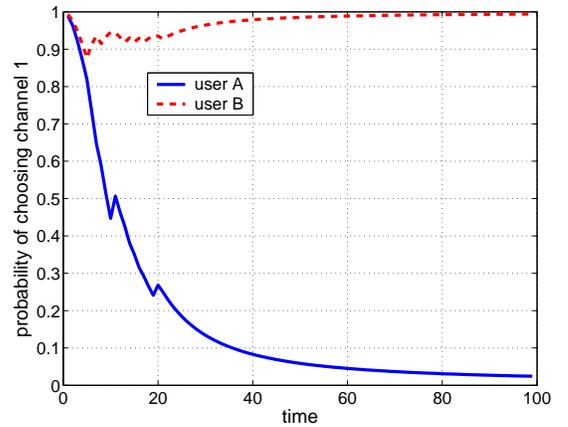}
  \caption{An example of the evolution of channel selection probability.}\label{fig:prob}
\end{figure}

Fig. \ref{fig:prob} shows the evolution of the probability of
choosing channel 1 when $\gamma=0.1$. We observe that both secondary
users prefer channel 1 at the beginning and soon secondary user $A$
intends to choose channel 2, thus avoiding the collision.

\subsection{Learning Speed}
\begin{figure}
  \centering
  \includegraphics[scale=0.4]{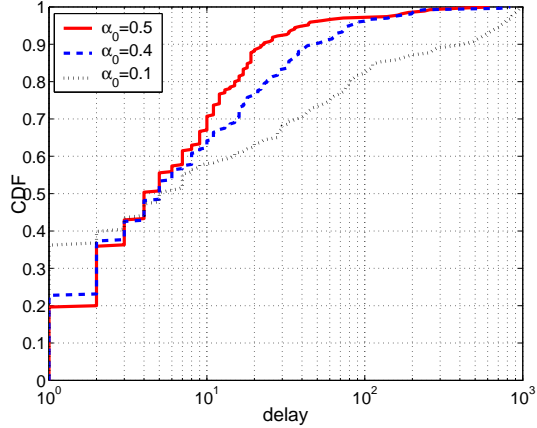}
  \caption{CDF of learning delay with different learning factor $\alpha_0$.}\label{fig:CDF1}
\end{figure}

\begin{figure}
  \centering
  \includegraphics[scale=0.4]{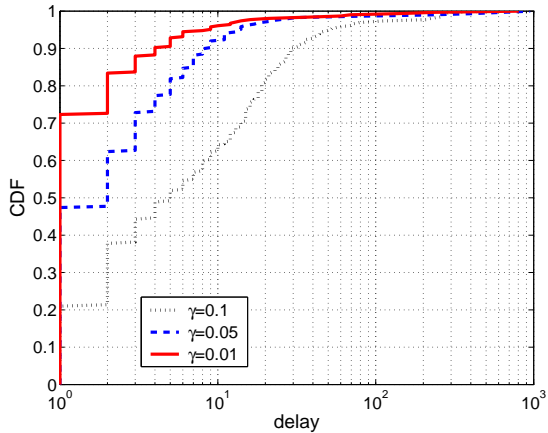}
  \caption{CDF of learning delay with different temperature $\gamma$.}\label{fig:CDF2}
\end{figure}

Figures \ref{fig:CDF1} and \ref{fig:CDF2} show the delays of
learning (equivalently, the learning speed) for different learning
factor $\alpha_0$ and different temperature $\gamma$, respectively.
The original $Q$ values are randomly selected. When the
probabilities of choosing channel 1 are larger than 0.95 for one
secondary user and smaller than 0.05 for the other secondary user,
we claim that the learning procedure is completed. We observe that
larger learning factor $\alpha_0$ results in smaller delay while
smaller $\gamma$ yields faster learning procedure.

\subsection{Fluctuation}
\begin{figure}
  \centering
  \includegraphics[scale=0.4]{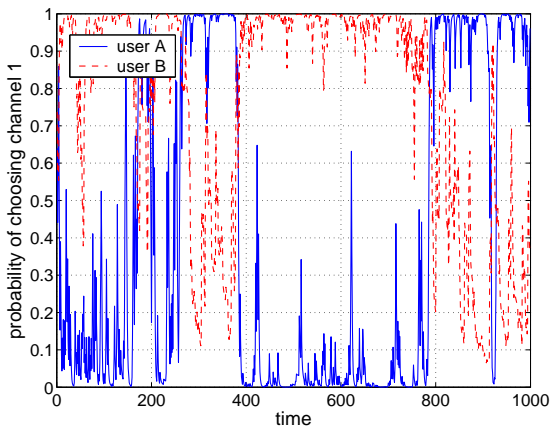}
  \caption{Fluctuation when improper lower bounds are selected.}\label{fig:fluc}
\end{figure}

In practical systems, we may not be able to use vanishing
$\alpha_{ij}(t)$ since the environment could change (e.g. new
secondary users emerge or the channel qualities change). Therefore,
we need to set a lower bound for $\alpha_{ij}(t)$. Similarly, we
also need to set a lower bound for the probability of exploring all
actions (notice that the exploration probability in
(\ref{eq:exploration}) can be arbitrarily small). Fig.
\ref{fig:fluc} shows that the learning procedure may yield
substantial fluctuation if the lower bounds are improperly chosen
(the lower bounds for $\alpha_{ij}(t)$ and exploration probability
are set as 0.4 and 0.2 in Fig. \ref{fig:fluc}).

\section{Conclusions}\label{sec:conclusion}
We have discussed the $2\times2$ case of learning procedure for
channel selection without negotiation in cognitive radio systems.
During the learning, each secondary user considers the channel and
the other secondary user as its environment, updates its $Q$ values
and takes the best action. An intuitive explanation for the
convergence of learning is provided using Metrick-Polak plot. By
applying the theory of stochastic approximation and ODE, we have
shown the convergence of learning under certain conditions.
Numerical results show that the secondary users can learn to avoid
collision quickly. However, if parameters are improperly chosen, the
learning procedure may yield substantial fluctuation.

\bibliographystyle{IEEE}

\end{document}